\documentclass[pre,showpacs]{revtex4}

\usepackage{amsmath}
\usepackage{amssymb}
\usepackage{graphicx}

\newcommand{\eps}{\varepsilon}

\newcommand{\prt}{\partial}

\begin{document}

\title{Whitham method for Benjamin-Ono-Burgers equation
and dispersive shocks in internal waves in deep fluid}

\author{Y. Matsuno$^{1}$}
\email{matsuno@yamaguchi-u.ac.jp}
\author{V.S. Shchesnovich$^{2}$}
\email{valery@loqnl.ufal.br}
\author{A.M. Kamchatnov$^{3}$}
\email{kamch@isan.troitsk.ru}
\author{R.A. Kraenkel$^{4}$}
\email{kraenkel@ift.unesp.br}

\affiliation{ $^1$Division of Applied Mathematical Science,
Graduate School of Science and
Engineering, Yamaguchi University,
Ube 755-8611, Japan\\
$^2$Instituto de F\'{\i}sica -
Universidade Federal de Alagoas, Macei\'o AL 57072-970, Brazil\\
$^3$Institute of Spectroscopy, Russian Academy of Sciences,
Troitsk 142190, Moscow Region, Russia\\
$^4$Instituto de F{i}sica Te\'{o}rica, Universidade Estadual
Paulista, Rua Pamplona 145, 01405-900 S\~{a}o Paulo, Brazil }

\begin{abstract}
The Whitham modulation equations for the parameters of a periodic
solution are derived using the generalized Lagrangian approach for
the case of damped Benjamin-Ono equation. The structure of the
dispersive shock in internal wave in deep water is considered
by this method.
\end{abstract}

\pacs{03.40.Kf, 03.40.Gc, 02.90.+p}

\maketitle

\section{Introduction}

As is known, dispersive shock is an oscillatory structure generated
in wave systems after wave breaking of intensive pulse under
condition that the dispersive effects are much greater than the
dissipative ones. In this sense, such shocks are dispersive
counterparts of usual viscous shocks well known in dynamics of
compressive viscous fluids. In the surface water waves physics, the
dispersive shocks are known as tidal bores in rivers \cite{BL}.
Besides this classical observation, the dispersive shocks have been
also found in some other physical systems including plasma
\cite{plasma} and Bose-Einstein condensate \cite{simula,hoefer}.

In typical situations, a dispersive shock can be represented as a
modulated nonlinear wave whose parameters change little in one
wavelength and one period; hence the Whitham modulation theory
\cite{whitham1,whitham2} (see also \cite{kamch2000}) can
be applied to its study. If one neglects dissipation, then
dispersive shock is a non-stationary structure expanding with time
so that at one its edge it can be represented as a soliton train and
at the other edge as a linear wave propagating with some group
velocity into the unperturbed region. A corresponding Whitham theory
of such shocks for the systems described by the Korteweg-de Vries
(KdV) equation was developed by Gurevich and Pitaevskii in \cite{GP1}
and later it was extended to other equations such as the
Kaup-Boussinesq system \cite{egp,egk1}, Benjamin-Ono (BO) equation
\cite{matsuno1,matsuno2,jorge},
and nonlinear Schr\"odinger equation \cite{kku}. This approach has
found applications to water waves physics \cite{apel} and dynamics of
Bose-Einstein condensate \cite{kgk1,hoefer}.

The Whitham method describes a long time evolution of the dispersive
shock when many waves (crests) are generated. However, when the
long-time evolution is considered, the small dissipation effects can
become of crucial importance. In particular, they can stop
self-similar expansion of the shock so that it tends to some
stationary wave structure which propagates as a whole with constant
velocity. Correspondingly, the Whitham equations should be modified
to include the dissipation effects. For the first time it was done
for the KdV-Burgers equation in \cite{gp2,akn} by direct
method which did not use applicability of the inverse scattering
transform method to the KdV equation. More general approach based on
the complete integrability of unperturbed wave equations was
developed in \cite{kamch04} and applied to the theory of bores described by
the Kaup-Boussinesq-Burgers equation \cite{egk2} and the KdV equation
with Chezy friction and a bottom with a slope \cite{egk3}.

The method of Ref.~\cite{kamch04} can be applied in principle to any wave
equation which is completely integrable in framework of the AKNS
method \cite{akns} and any perturbation depending on the wave variables
and their space derivatives. However, the important BO equation
describing internal waves in stratified deep water includes the
non-local dispersion term and therefore it cannot be considered by
the method of \cite{kamch04}. Although the Whitham theory for the BO
was discussed in \cite{matsuno1,matsuno2,jorge,dk}, its generalization to taking into
account small dispersion effects has not been developed yet. The aim
of this paper is to develop the Whitham theory for the
Benjamin-Ono-Burgers equation
\begin{equation}\label{2-1}
    u_t+uu_x+Hu_{xx}=\eps u_{xx},
\end{equation}
where
\begin{equation}\label{2-2}
    Hu(x)=\frac1{\pi}\mathrm{P.V.}\int\frac{u(y)}{y-x}dy
\end{equation}
is the Hilbert transform and the term in the right-hand side of
(\ref{2-1}) describes small friction with the viscosity parameter
$\epsilon$. In the next Section we shall derive the Whitham
equations which govern slow evolution of the nonlinear periodic wave
due to its modulation and small friction and in Section III we shall
apply this theory to the stationary bore (dispersive shock).

\section{Whitham theory for the Benjamin-Ono-Burgers equation}

The unperturbed BO equation has a periodic solution
\begin{equation}\label{3-1}
    u(x,t)=\frac{4k^2}{\sqrt{A^2+4k^2}-A\cos\theta}+\beta,\quad
    \theta=kx-\omega t,
\end{equation}
which depends on three constant parameters---wavenumber $k$,
amplitude of oscillations $A=(u_{max}-u_{min})/2$, and $\beta$. This
solution describes a nonlinear wave propagating with constant
velocity
\begin{equation}\label{3-2}
    V=\frac{\omega}k=\frac12\sqrt{A^2+4k^2}+\beta.
\end{equation}
In a modulated wave these three parameters become slow functions of
the space and time coordinates and their evolution is governed by
the Whitham equations which were obtained in \cite{dk} for the general
multi-phase solutions of the unperturbed BO equation by the method
based on the complete integrability of the BO equation and in
\cite{matsuno1,matsuno2} for the simplest one-phase solution (\ref{3-1}) by a direct
Whitham method based on the use of the Hamilton principle
\begin{equation}\label{3-3}
    \delta\int dt\int dx L(\phi, \phi_x,\phi_t)=0
\end{equation}
with the Lagrangian
\begin{equation}\label{3-4}
    L=\frac12\phi_t\phi_x+\frac16\phi_x^3+\frac12\phi_xH\phi_{xx},
    \quad u=\phi_x.
\end{equation}
In this method the periodic solution is represented in the form
\begin{equation}\label{3-5}
    \phi=\psi+\Phi(\theta),
\end{equation}
where
\begin{equation}\label{3-6}
    \psi=\beta x-\gamma t,\quad \theta=kx-\omega t,
\end{equation}
so that
\begin{equation}\label{3-7}
    \beta=\psi_x,\quad \gamma=-\psi_t,\quad k=\theta_x,\quad
    \omega=-\theta_t;
\end{equation}
hence $u=u(\theta,\theta_x,\theta_t,\psi_x)$,
$L=L(\theta,\theta_x,\theta_t,\psi_x,\psi_t)$, and the averaging is
taken over fast oscillations according to the rule
\begin{equation}\label{3-8}
    \overline{L}=\frac1{2\pi}\int_0^{2\pi} Ld\theta
\end{equation}
leading to the averaged Lagrangian which depends on the derivatives
$\theta_x,\theta_t,\psi_x,\psi_t$. The Euler-Lagrange equations for the
corresponding averaged Hamilton principle
\begin{equation}\label{4-1}
    \delta\int dt\int dx\overline{L}(\theta_x,\theta_t,\psi_x,\psi_t)=0
\end{equation}
yields with account of (\ref{3-7}) the Whitham equations in the form
\begin{equation}\label{4-2}
    \frac{\prt}{\prt t}\frac{\prt\overline{L}}{\prt\gamma}-
    \frac{\prt}{\prt x}\frac{\prt\overline{L}}{\prt\beta}=0,\quad
    \frac{\prt}{\prt t}\frac{\prt\overline{L}}{\prt\omega}-
    \frac{\prt}{\prt x}\frac{\prt\overline{L}}{\prt k}=0,
\end{equation}
which should be complemented by the consistency conditions
\begin{equation}\label{4-2a}
    \beta_x+\gamma_t=0,\quad k_t+\omega_x=0.
\end{equation}
After calculation of the     integral (\ref{3-8}) they reduce to
the system of equations for the parameters $\beta, k, V=\omega/k$
(see \cite{matsuno1,matsuno2}).

Now our task is to generalize this procedure to the perturbed BO equation
(\ref{2-1}). Instead of using the Lagrangian formulation with an additional field
(see, for instance, Ref.~\cite{KaupMalomed}) we prefer to introduce a more simple
approach which does not require introduction of new auxiliary fields. We propose to
use directly the Hamilton principle in its infinitesimal form by noticing that
Eq.~(\ref{2-1}) can be written symbolically as
\begin{equation}\label{4-3}
    \int dt\int dx\left\{ \delta L+\eps\phi_{xxx}\delta\phi\right\}=0,
\end{equation}
where $\delta L$ is a variation of the Lagrangian (\ref{3-4}).
Now we can transform (\ref{4-3}) in the following way. First, we
integrate the ``dissipative'' term by parts and use $u=\phi_x$:
\begin{equation}\label{4-4}
    \int dt\int dx\left\{ \delta L - \eps u_{x}\delta u\right\}=0.
\end{equation}
Second, we average the dissipative term as follows:
\begin{equation}\nonumber
    \begin{split}
    \int dx\overline{(u_x\delta u)}=\int dx\overline{[u_x(u_\theta\delta\theta
    +u_{\theta_x}\delta\theta_x+u_{\theta_t}\delta\theta_t+
    u_{\psi_x}\delta\psi_x)]}\\
    =\int dx\left\{ \overline{\left[ u_x u_\theta-
    \frac{\prt}{\prt x}(u_xu_{\theta_x})-
    \frac{\prt}{\prt t}(u_xu_{\theta_t})\right]}\delta\theta-
    \frac{\prt}{\prt x}\overline{(u_xu_{\psi_x})}\delta\psi \right\}.
    \end{split}
\end{equation}
In the Whitham approximation with fast $\theta$-variable we have
$\prt/\prt x\cong\theta_x\prt/\prt\theta$,
$\prt/\prt t\cong\theta_t\prt/\prt\theta$,
where within the averaging interval the parameters $\theta_x$ and
$\theta_t$ can be considered constant, the terms with $\theta$-derivatives
become equal to zero after averaging and, hence, we arrive at the expression
\begin{equation}\label{4-5}
    \int dt\int dx\overline{(u_x\delta u)}=\int dt\int dx \overline{(u_xu_\theta)}
    \delta\theta =\int dt\int dx k \overline{(u_\theta^2)}\delta\theta.
\end{equation}
Transformation of the term with Lagrangian in (\ref{4-4}) can be
performed in a similar way and as a result we obtain the Whitham
equations in the form
\begin{equation}\label{5-1}
    \frac{\prt}{\prt t}\frac{\prt\overline{L}}{\prt\gamma}-
    \frac{\prt}{\prt x}\frac{\prt\overline{L}}{\prt\beta}=0,\quad
    \frac{\prt}{\prt t}\frac{\prt\overline{L}}{\prt\omega}-
    \frac{\prt}{\prt x}\frac{\prt\overline{L}}{\prt k}=\eps k\overline{(u_\theta^2)},
\end{equation}
which generalize Eqs.~(\ref{4-2}) to the BOB equation (\ref{2-1}).

Simple calculation of averaged values gives \cite{matsuno1,matsuno2}
\begin{equation}\label{5-2}
    \overline{L}=\frac13k^3-k\left(\frac{\omega^2}{k^2}-\frac{\beta\omega}k+\gamma
    \right)+\frac16\beta^3-\frac12\beta\gamma,
\end{equation}
\begin{equation}\label{5-3}
    \eps k\overline{(u_\theta^2)}=2\eps(V-\beta)[(V-\beta)^2-k^2].
\end{equation}
Their substitution into (\ref{5-1}) and use of (\ref{4-2a}) permit one to express
$\gamma$ as $\gamma=\beta^2/2$ and transform equations for the other parameters
to the form
\begin{equation}\label{5-4}
\begin{split}
  &\beta_t+\beta\beta_x = 0, \\
  &k_t+(Vk)_x = 0, \\
  &V_t+VV_x+kk_x = -\eps(V-\beta)[(V-\beta)^2-k^2].
  \end{split}
\end{equation}
These are the Whitham equations for the physical parameters $\beta, k, V$.

Although equations (\ref{5-4}) are simple enough for further investigations,
they can be transformed to theoretically more attractive diagonal form by
introduction of Riemann invariants $a,b,c$ according to definitions
\begin{equation}\label{5-7}
    \beta=2c,\quad k=b-a,\quad V=b+a,\quad (c<a<b),
\end{equation}
so that we get the system
\begin{equation}\label{5-8}
    \begin{split}
    &a_t+2aa_x=-2\eps(a-c)(b-c)(a+b-2c),\\
    &b_t+2bb_x=-2\eps(a-c)(b-c)(a+b-2c),\\
    &c_t+2cc_x=0.
    \end{split}
\end{equation}
In terms of Riemann invariants the periodic solution (\ref{3-1}) takes the form
\begin{equation}\label{5-9}
    u(x,t)=\frac{2(b-a)^2}{a+b-2c-2\sqrt{(a-c)(b-c)}\,\cos\theta}+2c,
\end{equation}
where
\begin{equation}\label{5-10}
\theta=(b-a)x-(b^2-a^2)t.
\end{equation}
The parameters $a,b,c$ must satisfy the condition $2c<a+b$ to keep the solution
non-singular. The amplitude of oscillations is expressed as
\begin{equation}\label{5-11}
    A=4\sqrt{(a-c)(b-c)}.
\end{equation}

When for some concrete problem the solution of Eqs.~(\ref{5-8}) is found and
the functions $a=a(x,t),$ $b=b(x,t),$ $c=c(x,t)$ are known, their substitution into
(\ref{5-9}) yields the modulated nonlinear wave for the problem under consideration.
Mean value of $u$ in this oscillatory region is equal to
\begin{equation}\label{5-12}
    \overline{u}=2k+\beta=2(b+c-a).
\end{equation}
In the next Section we shall consider an example of such a problem.

\section{Dispersive shock (bore) in internal waves in a deep fluid}

As was mentioned in the Introduction, the dispersive shock is an oscillatory
region joining two regions with different values of the wave amplitude $u_\pm$
which arise after wave breaking. In the simplest case of the Gurevich-Pitaevskii
problem one can consider $u_\pm$ as two constants: $u\to u_\pm$ as $x\to\pm\infty$,
respectively, and without loss of generality we can take $u_+=0$ and denote
$u_-=u_0=\mathrm{const}$. Thus, we have to find the solution of the Whitham
equations which corresponds to a modulated nonlinear wave satisfying the
boundary conditions
\begin{equation}\label{6-1}
    u\to\left\{
    \begin{array}{c}
    u_0,\quad x\to-\infty,\\
    0,\quad x\to+\infty.
    \end{array}
    \right.
\end{equation}
Naturally, the initial profile of the wave $u=u(x,0)$ must also satisfy this
condition and easy estimate shows that after wave breaking the waves are generated
with wavelength $L\sim1/u_0$.

Now, we can distinguish two typical stages of evolution of the wave:
\begin{enumerate}
    \item The initial stage for $t\ll1/(\eps u_0^2)$, when we have
    $|u_t|\gg\eps|u_{xx}|$, so that one can neglect a viscous term in (\ref{2-1});
    \item The asymptotic stage of large time $t\gg 1/(\eps u_0^2)$, when the solution
    tends to the stationary solution determined by interplay of the dispersion and
    dissipation effects.
\end{enumerate}
Examples of the first stage have already been studied in \cite{matsuno2,jorge}.
The simplest case of a step-like initial condition has been discussed in \cite{matsuno1},
and we shall reproduce some results here for convenience of future comparison with the
second stage of asymptotically large time.

Thus, we suppose that at the initial moment $t=0$ the region of transition from
$u=u_0$ to $u=0$ is very narrow (i.e. much less than $L\sim 1/u_0$) so that the
initial profile can be approximated by a step-like function,
\begin{equation}\label{7-1}
    u(x,0)=\left\{
    \begin{array}{c}
    u_0,\quad x<0,\\
    0, \quad x>0.
    \end{array}
    \right.
\end{equation}
As was noticed above, at the initial stage we can neglect the dissipation effects,
so that arising dispersive shock is governed by the equations
\begin{equation}\label{7-2}
    a_t+2aa_x=0,\quad b_t+2bb_x=0,\quad c_t+2cc_x=0.
\end{equation}
After averaging over wavelength the initial-value problem for the Riemann invariants
does not contain any parameters with dimension of length. Hence, the Riemann
invariants can depend on the self-similar variable $\zeta=x/t$ only and the
Whitham equations (\ref{7-2}) reduce to
\begin{equation}\label{7-3}
    a_\zeta(2a-\zeta)=0,\quad b_\zeta(2b-\zeta)=0,\quad c_\zeta(2c-\zeta)=0.
\end{equation}
Since at one edge of the oscillatory region we must have $a=b$ (the soliton
limit with $k=0$) and at the other edge $a=c$ (the vanishing amplitude of
oscillations limit $A=0$), the only acceptable solution of these equations
reads $a=\zeta/2=x/2t$, $b=\mathrm{const},$ $c=\mathrm{const}$. According to
(\ref{7-1}), at the soliton limit $x=x_+$ $(a=b)$ the mean value $\overline{u}$
must match to the zero value at the right edge, $\overline{u}=2c=0$, hence
$c=0$ everywhere. At the opposite edge $x=x_-=0$ the mean value must
coincide with $u=u_0$, $\overline{u}=2b=u_0$, hence $b=u_0/2$ everywhere.
Thus, we arrive at the following solution of the Whitham equations (see Fig.~1)
\begin{equation}\label{8-1}
    \begin{split}
    &a={x}/{2t},\quad 0<x<x_+,\\
    &b={u_0}/2,\quad x<x_+,\\
    &c=0,\quad x>x_-=0,
    \end{split}
\end{equation}
where
\begin{equation}\label{8-2}
    x_+=u_0t.
\end{equation}
\begin{figure}[bt]
\includegraphics[width=8cm,height=5.5cm,clip]{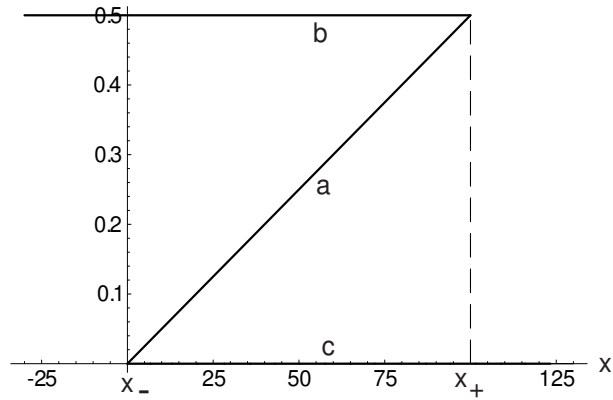}
\caption{Plots of the Riemann invariants at $t=100$ as functions of $x$
in the case of decay of the step-like initial distribution without friction.
} \label{fig1}
\end{figure}
Substitution of this solution into (\ref{5-9}) gives the expression for $u$
in the oscillatory region,
\begin{equation}\label{8-2a}
     u(x,t)=\frac{(u_0t-x)^2}{t(u_0t+x-2\sqrt{u_0tx}\cos\theta)},
\end{equation}
where
\begin{equation}\label{8-2b}
    \theta=[2(u_0t-x)x-(u_0^2t^2-x^2)]/(4t).
\end{equation}
The corresponding plot of the dispersive shock profile at fixed $t$ is shown
in Fig.~2.
\begin{figure}[bt]
\includegraphics[width=8cm,height=5.5cm,clip]{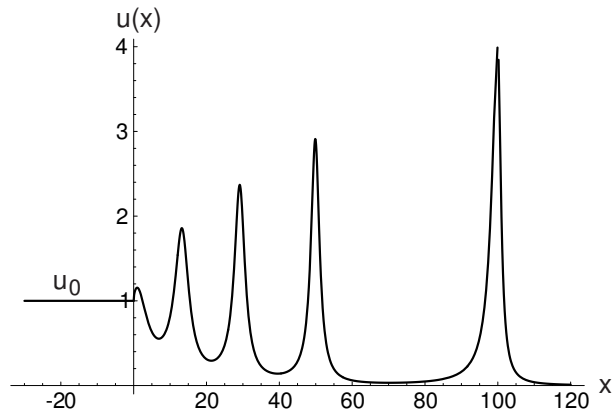}
\caption{Profile of the bore at $t=100$ corresponding to the self-similar
decay of the step-like initial distribution without friction.
} \label{fig2}
\end{figure}
At the leading front we can see a soliton with the amplitude
\begin{equation}\label{8-3}
    u_s=4V=4u_0
\end{equation}
which moves to the right with velocity $V=u_0$. The trailing edge is located
at $x=0$ and corresponds to a linear wave with vanishing amplitude and zero
value of group velocity. Indeed, linearization of the BO equation with respect
to small amplitude $A$ in $u\cong U_0+A\cos(kx-\omega t)$ leads to the
dispersion relation
\begin{equation}\label{8-3a}
    \omega=u_0k-k^2.
\end{equation}
According to (\ref{5-7}) we have $k=b=u_0/2$ at $x=0$ and hence
\begin{equation}\label{8-3b}
    \left.\frac{d\omega}{dk}\right|_{k=u_0/2}=0\quad\text{at}\quad x=0.
\end{equation}
Number of waves in the oscillatory region is equal to
\begin{equation}\label{8-3c}
    N=\frac1{2\pi}\int_0^{u_0t}kdx=\frac{u_0^2t}{8\pi}.
\end{equation}

The self-similar expansion of the oscillatory region holds as long as the
viscosity effects can be neglected. However, these effects come into play
at $t\sim 1/(\eps u_0^2)$ and at $t\to\infty$ the shock profile tends to the
stationary structure propagating with constant velocity. To find this
structure, we look for the stationary solution of the Whitham equations
(\ref{5-8}) so that the Riemann invariants are functions of $\xi=x-Vt$
only, where $V=a+b=\mathrm{const}$. We assume that $\xi=0$ corresponds to
the leading soliton front of the shock, where $k=b-a=0$ and $\overline{u}=
2(b+c-a)=0$, which give at once that $c=0$ and $a=b=V/2$ at $\xi=0$.
Then the last equation (\ref{5-8}) gives $c=0$ identically and the rest
equations (\ref{5-8}) reduce to a single equation
\begin{equation}\label{8-4}
    (2b-V)b_\xi=-2\eps Vb(V-b)
\end{equation}
which should be solved with the initial condition
\begin{equation}\label{8-5}
    \left. b\right|_{\xi=0}=\frac{V}2.
\end{equation}
Elementary calculation with account of inequality $b>a=V-b$, i.e.
$b\geq V/2$, gives at once $b=(V/2)[1+\sqrt{1-\exp(2\eps V\xi)}]$.
At last, at $\xi\to-\infty$ we must have $a=0$, $b=V$, $\overline{u}
=u_0=2b=2V$, that is $V=u_0/2$, and we arrive at the solution
\begin{equation}\label{8-6}
    a=\frac{u_0}4\left[1-\sqrt{1-\exp(2\eps V\xi)}\right],\quad
    b=\frac{u_0}4\left[1+\sqrt{1-\exp(2\eps V\xi)}\right],\quad
    c=0,
\end{equation}
where $\xi=x-(u_0/2)t$. Plots of the Riemann invariants are shown in
Fig.~3; they should be compared with Fig.~1.
\begin{figure}[bt]
\includegraphics[width=8cm,height=5.5cm,clip]{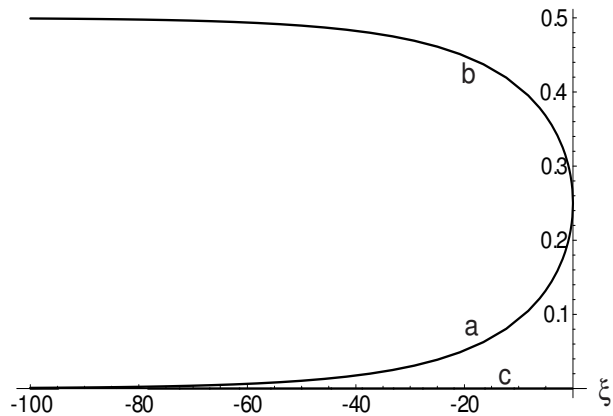}
\caption{Plots of the Riemann invariants as functions of $\xi$
in the case of stationary bore solution with friction ($\eps=0.05$).
} \label{fig3}
\end{figure}
 Substitution of
Eq.~(\ref{8-6}) into Eq.~(\ref{5-9}) yields the profile of the
shock,
\begin{equation}\label{9-1}
    u(x,t)=\frac{u_0[1-\exp(\eps u_0\xi)]}
    {1-\exp(\eps u_0\xi/2)\,\cos\theta}
\end{equation}
where
\begin{equation}\label{9-2}
    \theta=\frac{u_0}2\sqrt{1-\exp(\eps u_0\xi)}\,\xi,\quad
    \xi=x-\frac{u_0}2t.
\end{equation}
The corresponding plot is shown in Fig.~4.
\begin{figure}[bt]
\includegraphics[width=8cm,height=5.5cm,clip]{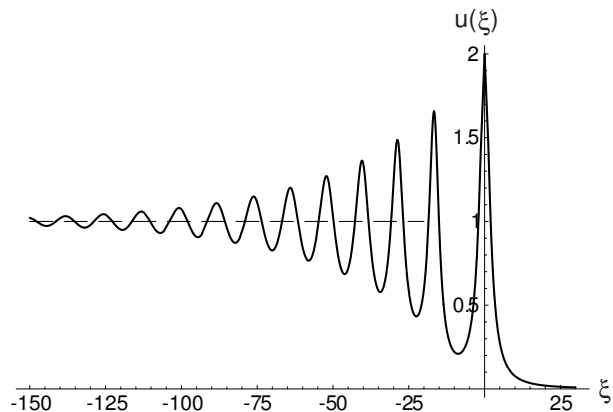}
\caption{Profile of the stationary bore described by the
Bebjamin-Ono-Burgers equation with friction coefficient
$\eps=0.05$. Dashed line corresponds to the constant
value of $u$ at $x\to-\infty$.
} \label{fig4}
\end{figure}
Again at the leading
front we can see a soliton, but now it has the amplitude
\begin{equation}\label{9-3}
    u_s=2u_0
\end{equation}
and propagates with velocity $V=u_0/2$. Thus, the friction effects
lead to decrease of the amplitude and velocity of the soliton
compared with the non-stationary stage. However, it is important
to notice that, on the contrary to an isolated soliton, the profile
of the bore becomes asymptotically stationary. In this stationary
solution the trailing edge is located at $x=-\infty$.

\section{Conclusion}

In this paper, we have discussed the structure of dispersive shock
in the internal wave subject to small friction effects. It is shown
that there exists a stationary profile so that non-stationary
oscillating structure supported by a jump of the wave amplitudes
at two spatial infinities tends asymptotically to this stationary
profile.

The Whitham method applied to this problem occurs very simple in the
case of BOB equation and the corresponding equations can be solved in
elementary and explicit form.

\section*{Acknowledgements}

The work of V.S.S. was supported by the CNPq grant. A.M.K. thanks
FAPESP for support of his stay at IFT-UNESP, Brazil. V.S.S. thanks
the IFT-UNESP for warm hospitality and financial support during his
visit.

\end{document}